\documentclass[journal=jacsat,manuscript=article]{achemso}
\usepackage[version=3]{mhchem} 
\usepackage[]{graphicx}
\usepackage{amsmath}
\usepackage{times}
\usepackage{units}
\usepackage{color}
\usepackage{mathtools}
\usepackage{amsfonts}
\usepackage{lineno}

\usepackage{natbib}

\author{Lucas Lafeta}
\affiliation{Department of Chemistry and CeNS, LMU Munich, Butenandtstr. 5-13, 81377 Munich, Germany.}
\email{lucas.lafeta@cup.lmu.de}
\author{Sean Hartmann}
\affiliation{Department of Chemistry and CeNS, LMU Munich, Butenandtstr. 5-13, 81377 Munich, Germany.}
\author{Bárbara Rosa}
\affiliation{Institute of Solid State Physics, Technische Universität Berlin, Hardenbergstraße 36, 10623 Berlin, Germany.}
\author{Stephan Reitzenstein}
\affiliation{Institute of Solid State Physics, Technische Universität Berlin, Hardenbergstraße 36, 10623 Berlin, Germany.}
\author{Leandro M. Malard}
\affiliation{Departamento de Física, Universidade Federal de Minas Gerais, Av. Antônio Carlos 6627, Belo Horizonte, Minas Gerais 30123-970, Brazil.}
\email{lmalard@fisica.ufmg.br}
\author{Achim Hartschuh}
\affiliation{Department of Chemistry and CeNS, LMU Munich, Butenandtstr. 5-13, 81377 Munich, Germany.}
\email{achim.hartschuh@lmu.de}

\title{Probing Noncentrosymmetric 2D Materials by Fourier Space Second Harmonic Imaging}

\begin{document}


\begin{abstract}

The controlled assembly of twisted 2D structures 
requires precise determination of the crystal orientation of their component layers. 
In the established procedure, the second-harmonic generation (SHG) intensity of a noncentrosymmetric layer is recorded while rotating the polarization of both 
the incident laser field and detected SHG, which can be time-consuming and tedious. 
Here, we demonstrate that the crystal orientation of
transition metal dichalcogenides and hexagonal boron nitride can be directly determined 
by recording SHG images generated by tightly focused laser beams in Fourier space. 
Using an azimuthally polarized laser beam, the 
SHG image distinctly reflects the hexagonal structure of the crystal lattice, revealing its 
orientation quickly and accurately.
This technique could significantly impact the field of twistronics, 
which studies the effects of the relative angle between the layers 
of a stacked 2D structure, as well as advance the nanofabrication of 2D materials.

\end{abstract}

The field of two-dimensional (2D) crystalline materials has seen a dramatic development within the last two decades. Apart from graphene, which is considered the first 2D material to be isolated, a vast number of other 2D materials exist, and this number continues expanding. In addition to their intriguing physical properties, these layered materials are ideally suited for forming stacked van der Waals (vdW) structures. The electro-optical properties of these structures are significantly influenced by the relative angle (twist) between the layers. This variable has proven crucial for numerous angle-dependent physical phenomena, leading to the emergence of the fields of twist-optics and twistronics\cite{li_lattice_2021, yankowitz_tuning_2019, ciarrocchi_excitonic_2022, forg_moire_2021, arora_superconductivity_2020, tarnopolsky_origin_2019, carr_twistronics_2017, devakul_magic_2021, shabani_deep_2021, wu_topological_2019, andersen_excitons_2021, sung_broken_2020, weston_atomic_2020, enaldiev_stacking_2020, scuri_electrically_2020, xu_tunable_2022, li_observation_2010, gadelha_localization_2021}. 

A key experimental challenge in this field is determining the crystal orientation of individual flakes \cite{mak_semiconductor_2022, kapfer_programming_2023, wang_clean_2023, liao_precise_2020} while developing practical, fast, and precise techniques to achieve this.
A widely used procedure for noncentrosymmetric materials is based on 
polarization-resolved detection of second-harmonic generation (SHG) \cite{psilodimitrakopoulos_ultrahigh-resolution_2018, autere_nonlinear_2018, li_probing_2013, ma_rich_2020}.
This technique exploits the particular symmetry properties of the investigated materials and can be readily applied to a broad range of materials \cite{boyd_chapter_2008,shen_optical_1989,abdelwahab_giant_2022,lucking_large_2018,shree_interlayer_2021,seyler_electrical_2015,carvalho_nonlinear_2020,malard_observation_2013,mennel_second_2018,zuo_optical_2020}. 

Within the 2D materials family semiconducting transition metal dichalcogenides (TMDs) have attracted particular interest because of their strong linear and nonlinear optical response, which stems from the large oscillator strength of tightly bound excitons \cite{wang2018colloquiumReview,mueller2018review, malard_observation_2013,li_probing_2013,lafeta_second-_2021, kumar_second_2013}. For this 2D materials class, SHG detection turned out to be a crucial and powerful tool for determining the number of layers in a given flake and its crystallographic orientation \cite{malard_observation_2013,li_probing_2013}. 
Materials with a hexagonal crystalline structure, such as TMD monolayers (e.g.~MoSe$_2$, WSe$_2$) and hexagonal boron nitride (h-BN), belong to the symmetry point group D$_{3h}$, for which only three of the 27 components of the second-order susceptibility tensor $\chi^{(2)}$ that determines the SHG response are nonzero \cite{pike_angular_2022,boyd_chapter_2008,li_probing_2013}. This crystalline symmetry translates into the following polarization dependence: If the input polarization matches the zigzag direction of the TMD (Fig.~\ref{fig1}), the polarization of the emitted SHG is rotated by 90°. In contrast, input and output polarization directions are the same for input polarization parallel to the armchair direction. The crystal orientation is thus encoded in the polarization of the generated second harmonic response.
In the established and broadly applied procedure for determining the crystal symmetries and crystallographic direction of 2D materials, the SHG intensity is recorded while rotating the polarization of the incident laser field, together with polarization filtering of the generated SHG with respect to the fixed crystal orientation\cite{malard_observation_2013,li_probing_2013}. 
Being a sequential and, hence, a time-consuming experiment may limit its use for fast orientational monitoring in applications such as multi-component 2D material assembly.

Here, we investigated SHG radiation patterns of 2D materials with D$_{3h}$ symmetry in Fourier space. 
We find that in the case of a tightly focused Gaussian laser beam, the crystal orientation of the flakes can be derived from the ellipticity of the observed SHG patterns.  
A microscopic model that treats the SHG of the 2D material as a coherent superposition of the fields radiated by dipolar emitters within the area of the laser-illuminated layer and that considers the form of the $\chi^{(2)}$ tensor can quantitatively describe the detected patterns and their orientation dependence. 
For an azimuthally polarized laser beam in which the excitation polarization varies spatially, the detected SHG pattern directly reveals the hexagonal shape of the crystal lattice together with its orientation. 
The concept underlying this procedure is to encode polarization information into spatial information and leverage SHG's coherent nature.
The approach presented provides a fast and precise tool for determining the crystal symmetry and orientation of a 2D material flake, a critical challenge in vdW multilayer structure assembly.

\section{Fourier space SHG images of 2D materials}

\begin{figure}[htp]
\begin{center}
 \includegraphics[width=.5\columnwidth]{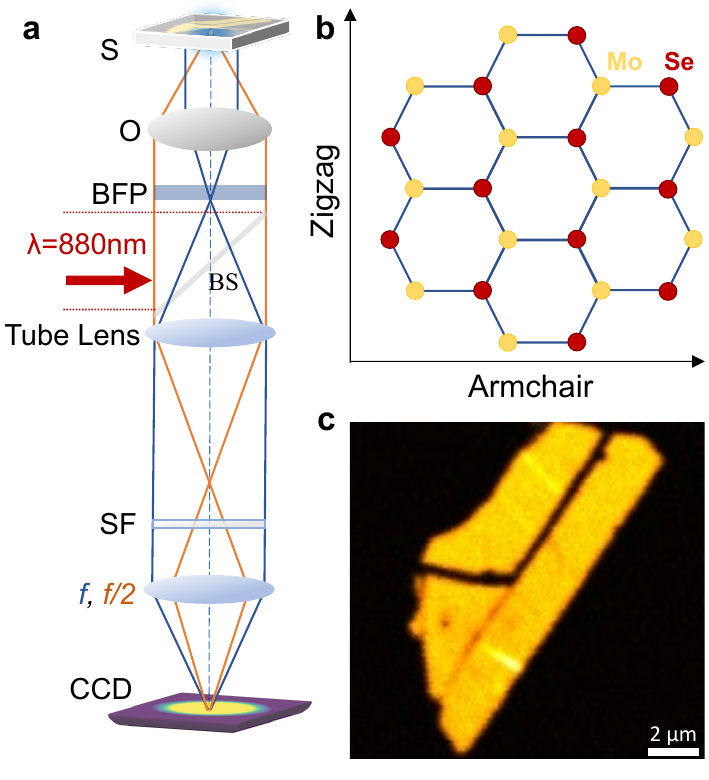}
\end{center}
 \caption{{\bf SHG imaging of 2D materials.} {\bf a} Microscope setup for the detection of real space and Fourier space (back focal plane, BFP) images. The fs-pulsed laser operating at 880 nm is directed to the objective (O) by a beam splitter (BS). The laser is focused on the sample, and the generated SHG at 440 nm is collected with the same objective, passing through a tube lens and a short pass filter (SF) to suppress the laser fundamental and detected by a CCD camera. Back focal plane imaging is done by placing a lens (known as Bertrand or Fourier lens) of focal length $f$ (blue path), and real space imaging is done by a $f/2$ focal length lens (orange path). {\bf b} Crystal structure of MoSe$_{2}$ 2D monolayers with  D$_{3h}$ symmetry. {\bf c} Confocal SHG scan image of a MoSe$_2$ flake.
}
\label{fig1} 
\end{figure}

In this study, SHG from 2D materials on glass substrates is detected either in real or Fourier space. The setup is based on a confocal microscope with a high numerical aperture objective (NA = 1.3) (Fig.~\ref{fig1}a, Supplementary Information Note 1). SHG images at 440 nm are recorded by a charge-coupled device (CCD) upon pulsed laser excitation at 880 nm. 
Fourier-space images that show the SHG radiation pattern, i.e., the angular distribution of SHG emission, are recorded in a conjugate back focal plane (BFP). Real-space images can be obtained by changing the focal length of the final focusing lens from f to f/2.
Fig.~\ref{fig1}b displays a representation of the crystallographic structure of a 2D monolayer with D$_{3h}$ symmetry indicating zigzag and armchair directions.
In the experiment, monolayer flakes are localized and identified by raster scanning the sample while detecting the SHG light using an avalanche photodiode. A representative confocal SHG image of a MoSe$_2$ flake obtained in this way is included in Fig.~\ref{fig1}c. The observed uniform SHG intensity confirms a spatially homogeneous monolayer flake. The general characterization of the samples is shown in Supplementary Information Note 2.

\begin{figure}[htp]
\begin{center}
\includegraphics[width=1\columnwidth]{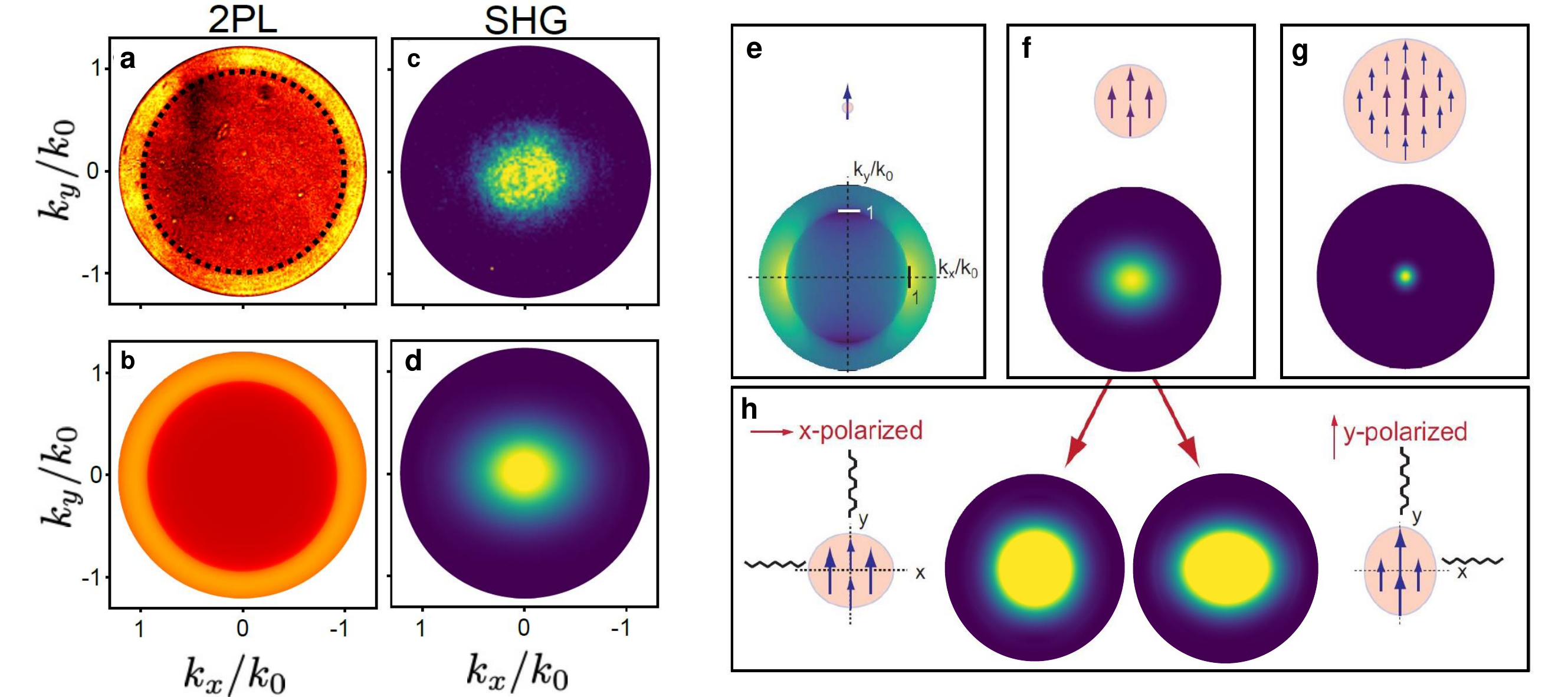}
\end{center}
 \caption{{\bf Origin of the ellipticity of SHG radiation patterns}. Experimental radiation patterns generated by 2PL in {\bf a} and SHG in {\bf c} were detected in the BFP. The dashed circle in {\bf a} denotes $(k_x/k_0)^2+(k_y/k_0)^2=1$. {\bf b} Calculated radiation pattern from the sum of incoherent point dipoles with isotropic in-plane orientation. {\bf d} Calculated pattern for coherent point dipoles oriented in the x-direction (see text and Methods section). {\bf e - g} Calculated radiation patterns generated by an increasing number of coherent point dipoles oriented in the x-direction. The red-shaded circles represent the laser-illuminated focal area. The radiation pattern of a single point dipole (left) features characteristic emission lobes for $(k_x/k_0)^2+(k_y^2/k_0) > 1$. No oriental information is contained in the circular pattern for weak focusing (right). For a tightly focused Gaussian laser beam (middle) with full-width at half maximum (FWHM) = 405 nm, information on the dipole orientation is retained in the form of an elliptical emission pattern. {\bf h} For both incident polarization parallel to the zigzag axis (x-polarized) and parallel to the armchair axis (y-polarized), the resulting SHG emission is polarized in y-direction. A tightly focused Gaussian laser beam features an elliptical intensity distribution with the long axis parallel to the laser polarization (red-shaded ellipses). This reduces the ellipticity of the SHG pattern for incident polarization parallel to the zigzag axis (left pattern) and enhances it for armchair polarization (right pattern). Note that the radiation patterns are observed in Fourier space, such that the long and short axis of the red-shaded areas appear inverted. The color scale was saturated for better visualization of ellipticity.} 

\label{fig2} 
\end{figure}

To understand the image contrast generated by nonlinear processes in 2D materials in Fourier space, we compare the results obtained for SHG with those for 2-photon excited photoluminescence (2PL), a nonlinear process that involves the simultaneous absorption of two photons \cite{2PL_1,2pl_2}. 
Fig.~\ref{fig2} presents the radiation patterns detected in Fourier space generated by 2PL and SHG obtained for a tightly focused  Gaussian laser beam.
The 2PL radiation pattern is radially symmetric and features the highest intensities for angles above the critical angle $\theta_{crit}$ 
corresponding to $(k_x / k_0)^2+(k_y / k_0)^2 \ge 1$ (Fig.~\ref{fig2}a), where $k_{x,y} / k_0$ denote the in-plane wavevector components normalized by the wavevector in vacuum $k_0$. The critical angle is determined by the ratio of the refractive indices of the glass/air interface according to $\sin \theta_{crit}=n_1 / n_2$. Because 2PL is a spontaneous process, the 2PL radiation pattern can be represented by the sum of the intensities of the radiation patterns generated by incoherent point dipolar emitters excited within the illuminated sample area. 
Using the model described in Supplementary Information Note 3, the calculated pattern presented in Fig.~\ref{fig2}b matches the experimental one without a free parameter.
The observed radial symmetry indicates that the 2PL from MoSe$_2$ at room temperature is unpolarized. This can be accounted for in the model 
calculations using orthogonal in-plane dipoles~\cite{budde_raman_2016}. 
In contrast, the SHG radiation pattern in Fig.~\ref{fig2}c is most intense for $k_{x,y} =0$ and slightly elliptical. Being a coherent process, SHG can be modeled as the sum of fields emitted by coherent dipoles, as shown in Fig.~\ref{fig2}d, where a small ellipticity can be seen as well. In the model calculation, the strength of the dipolar emitters placed at different positions within the flake is scaled with the spatial intensity and polarization distribution in the exciting laser focus 
(refer to Methods section and Supplementary Information Note 3).

Observing the ellipticity in both the experimental and calculated SHG radiation patterns in Fig.~\ref{fig2}, we explored its origin in detail starting with model calculations. For a single point dipole at a dielectric interface, the emission pattern directly reflects its orientation, with the maximum emission intensity occurring at angles greater than the critical angle (Fig.~\ref{fig2}e) \cite{lieb_single-molecule_2004}. 
For an extended illuminated area, on the other hand, the emitted SHG light can be calculated as a coherent superposition of the fields radiated by a spatial distribution of dipoles (Fig.~\ref{fig2}g)\cite{carvalho_nonlinear_2020}. 
In this large area limit, i.e.~with area dimensions exceeding the wavelength of the emitted light, the radiation pattern becomes a radially symmetric peak due to destructive interference of large angle contributions. 
Importantly, the resulting pattern contains no information on the orientation of the emitting dipoles.
For a tightly focused laser beam and a correspondingly narrow distribution of emitting dipoles, however, the calculated radiation pattern 
retains polarization information in the form of an ellipse oriented perpendicular to the direction of the emitting dipoles \cite{spychala_spatially_2020} (Fig.~\ref{fig2}f).
Although the orientation of this ellipse would enable us to determine the polarization of the emitted SHG light directly from Fourier images
and thus infer the crystal orientation, a secondary contribution to the ellipticity of the detected pattern arises, which can partially offset the first.

For a tightly focused x-polarized Gaussian laser mode, the distribution of the x-component of the electric field is elliptical due to the different projections of the p- and s-polarized fields onto the x-y-plane (Fig.~\ref{fig2}h) \cite{hecht_propagation_2012}. Hence, the in-plane intensity distribution of the generated SHG light in real space becomes elliptical, which translates into an additional contribution to the ellipticity of the radiation pattern that follows the polarization of the incident field.
As a result, the ellipticity of the pattern obtained for laser polarization parallel to the armchair axis (y-polarized in Fig.~\ref{fig2}h) is increased compared to that for laser polarization parallel to the zigzag axis. 
For polarization angles between x-and y-polarization, both ellipticity and the orientation of the ellipse vary gradually.
In summary, the ellipse's orientation and ellipticity result from the interplay between the radiation pattern of a narrow emitting
focal area, the nonlinear susceptibility tensor of noncentrosymmetric 2D materials that determine the polarization of the emitted SHG
light, and the field distribution of a tightly focused laser beam.

\section{Crystal orientation from Fourier SHG patterns}

\begin{figure}[htp]
\begin{center}
\includegraphics[width=1\columnwidth]{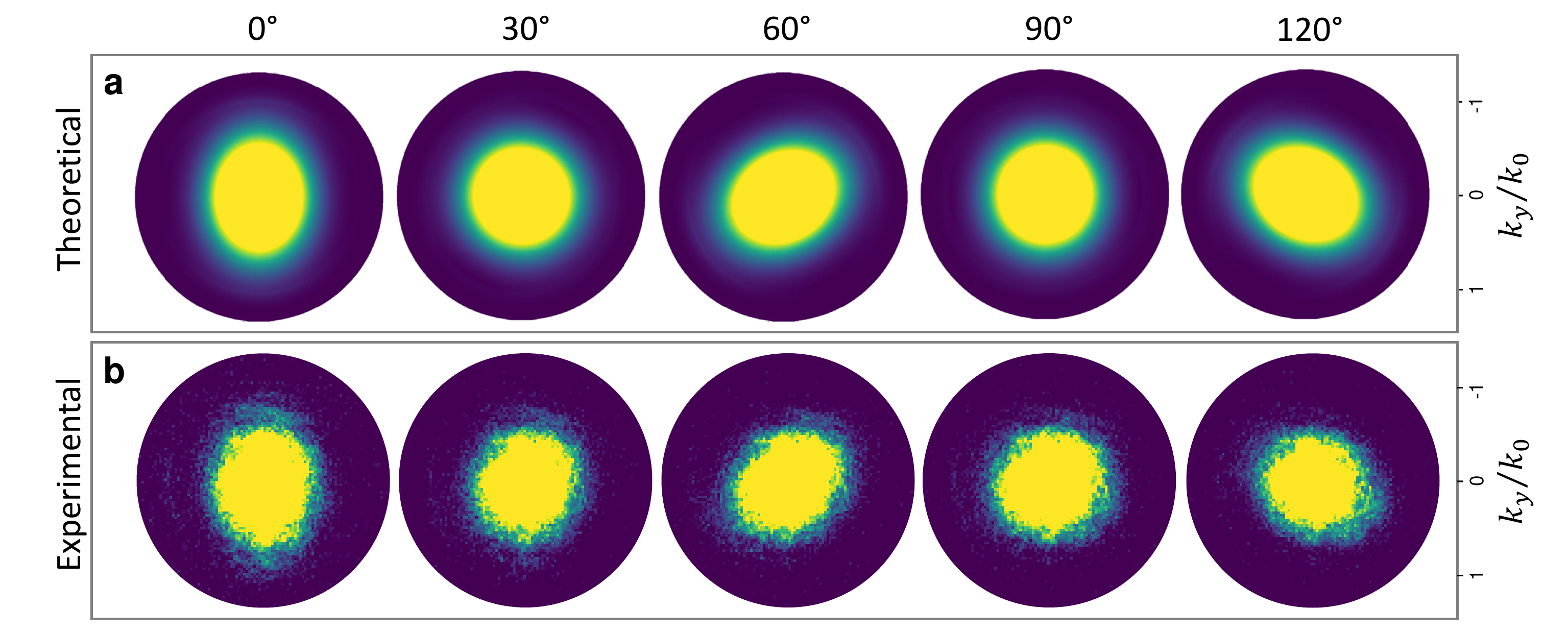}
\end{center}
 \caption{{\bf Excitation polarization dependence of SHG radiation patterns.}
{\bf a} Calculated SHG radiation patterns for D$_{3h}$ symmetry upon increasing the angle of the incident laser polarization with respect to the fixed directions of the crystallographic axis.
{\bf b} Experimental SHG radiation patterns for the corresponding incident laser polarizations using MoSe$_2$ monolayer.
}
\label{fig3} 

\end{figure}

To validate the model predictions for the polarization angle-dependent Fourier SHG patterns made above, we measured the Fourier SHG 
images generated by a MoSe$_2$ monolayer with known orientation while rotating the polarization of the incident Gaussian laser beam.
As can be seen in Fig.~\ref{fig3} for images taken every 30° both ellipticity and ellipse orientation vary in good agreement with the calculated patterns using the coherent dipole model described in the Methods section. 
The experimentally derived ellipticities are shown below in Fig.~\ref{fig5}b together with the calculated ones and the 6-fold intensity pattern using the established technique based on synchronously rotating incident laser polarization and detection polarization of the SHG light~\cite{li_probing_2013} (Fig.~\ref{fig5}a).
We can see that the ellipticity indeed follows the same dependence on input polarization. 
Importantly, the quantitative agreement with the calculated values shows that the ellipticity provides a direct measure of the crystal orientation in contrast to the intensity-based technique which involves sequential measurements.  
However, the ellipticity-based technique requires careful optical alignment and data analysis. 
A more robust and direct approach is presented in the next section using an azimuthally polarized laser mode.

\section{Fourier SHG images using azimuthally polarized laser mode}

\begin{figure}[htp]
 \begin{center}
 \includegraphics[width=0.9\columnwidth]{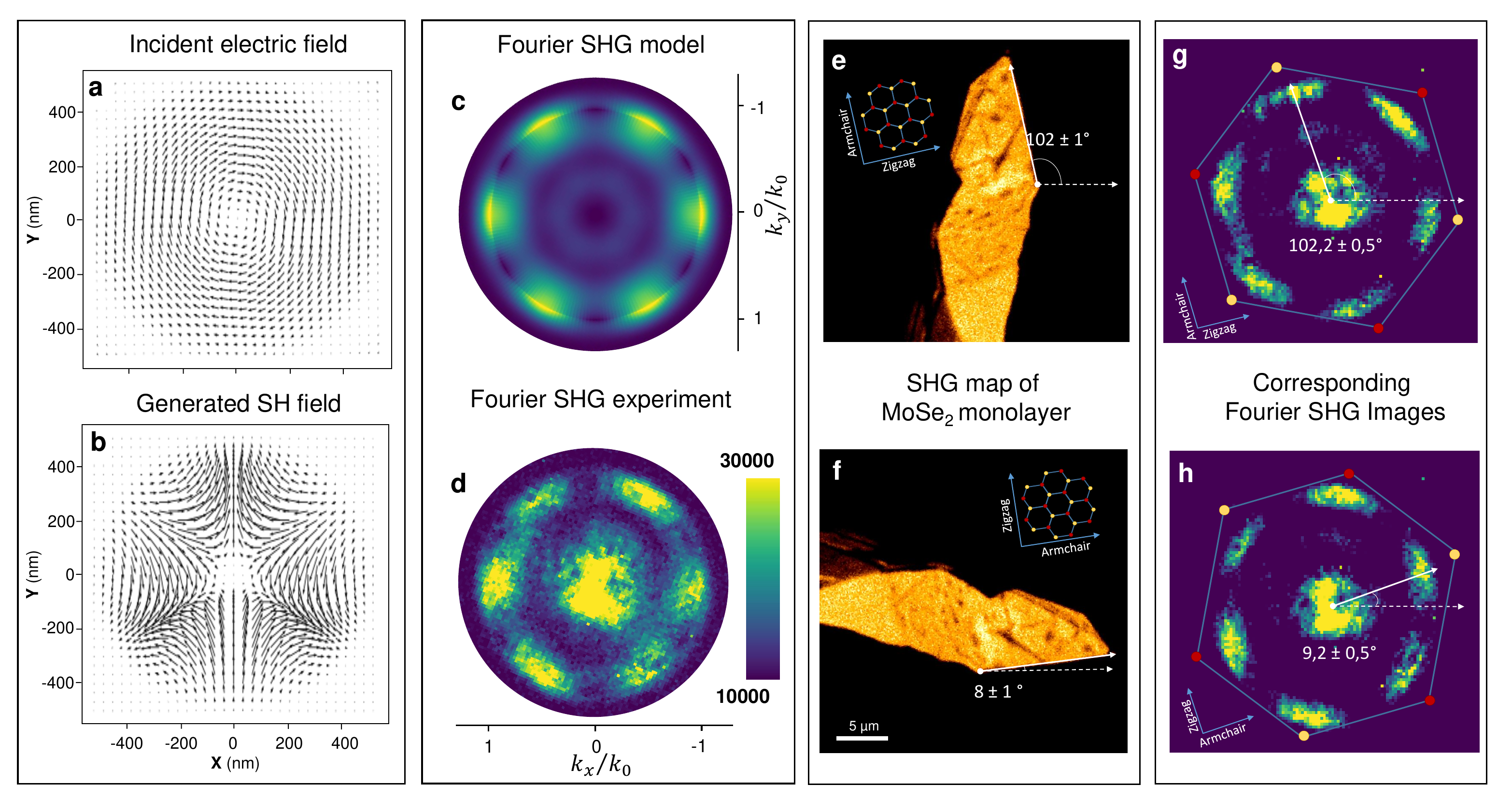}
 \end{center}
 \caption{{\bf Fourier SHG images for azimuthal mode excitation show
 the crystallographic orientation.} {\bf a} Focal field distribution of the incident azimuthal polarization mode. {\bf b} Focal field distribution of the SH generated by the D$_3h$ symmetry material. Fourier SHG images using azimuthal mode {\bf c} model and {\bf d} experiment, revealing the hexagonal structure of the material. 
 {\bf e-h} Confocal SHG and Fourier SHG images of a MoSe$_2$ flake before ({\bf e, g})  and after rotation by $\sim$90° ({\bf f, h}), respectively.    
 }
\label{fig4} 
\end{figure}

In the previous section, we derived information on the polarization direction of the SHG light by carefully studying its influence on the observed Fourier SHG images. Now, we follow a different approach and encode polarization information into the spatial coordinates of the exciting laser field in the form of an azimuthally polarized laser beam (Fig.~\ref{fig4}a). Depending on the local orientation of the input focal field relative to the armchair and zigzag directions, respectively, the SHG retains the polarization direction of the excitation light or appears rotated by 90°. The resulting field distribution of the SHG light reflects this complex interplay (Fig.~\ref{fig4}b). This field distribution then translates into the observed radiation pattern. Fig.~\ref{fig4}c presents the simulation results for the Fourier SHG image, and Fig.~\ref{fig4}d the experimental result for a MoSe$_2$ monolayer. Both patterns show the highest intensities around the critical angle (NA=1) and feature the 6-fold symmetry of the underlying crystal lattice.

\begin{figure}[htp]
\begin{center}
 \includegraphics[width=1\columnwidth]{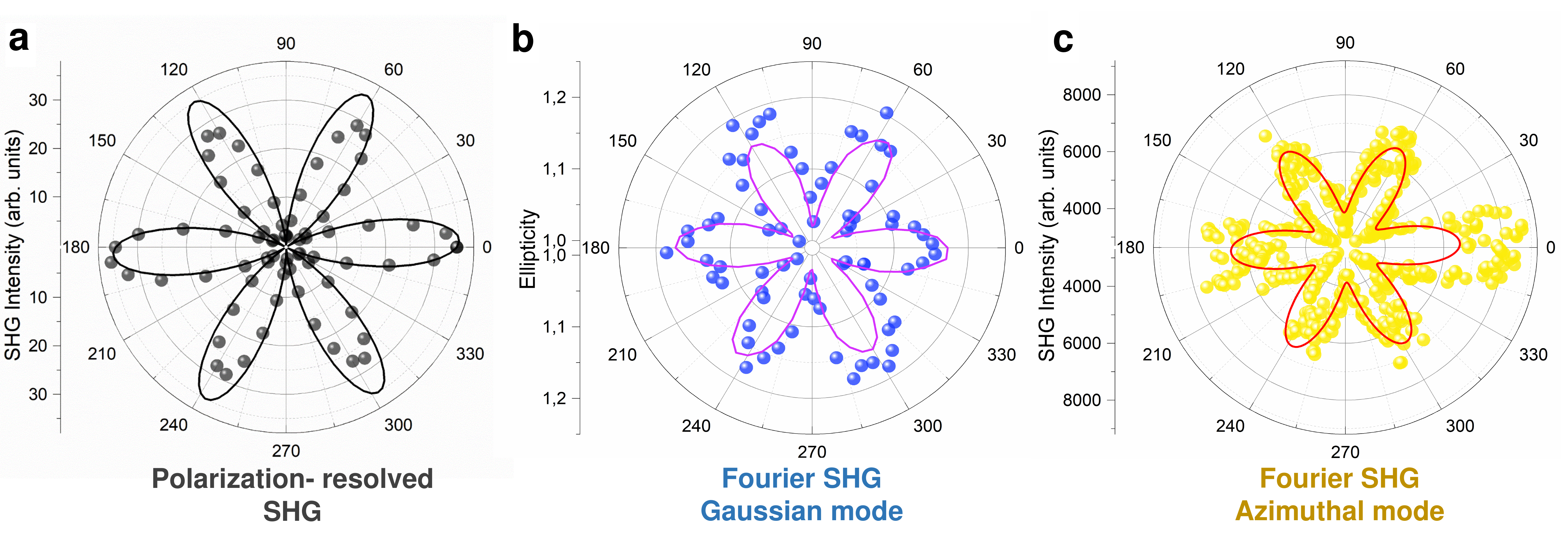}
\end{center}
 \caption{{\bf Comparison of techniques for determining crystal orientation.} {\bf a} SHG intensity as a function of the angle between the crystal armchair axis and the incoming and outgoing light polarization for a MoSe$_2$ monolayer sample. 
 The circles are the experimental data, while the black line is the theoretical six-fold SHG intensity pattern. 
  {\bf b} Ellipticity of the detected SHG radiation pattern as a function of incoming light polarization relative to the armchair axis. The blue circles are extracted from the experimental data, while the violet line is from the model calculations. 
  {\bf c} Plot polar of the intensity of Fourier SHG pattern using the azimuthal mode. The red line is the fit using the equation of Supplementary Information Note 4. The yellow circles are the intensity of the SHG extracted along the red circle of the MoSe$_2$ monolayer in Fig.~\ref{fig6} below.
}
\label{fig5} 

\end{figure}

To further confirm the correlation between the SHG radiation pattern and the crystal lattice orientation, we recorded confocal SHG images of the MoSe$_2$ flake to determine its orientation before and after rotation, using the observable edges of the flake as reference. Fig.~\ref{fig4}e and f show the flake at an initial orientation of 102° with respect to the horizontal and after rotation at 8°, respectively. The corresponding radiation patterns in Fig.~\ref{fig4} g, h reflect this rotation.
As demonstrated in Supplementary Information Note 4, Fourier SHG detection using an azimuthal laser beam enables the discrimination of rotation angles smaller than 
$\sim$0.5° with SHG image acquisition times of 1 second.
In Fig.~\ref{fig5}c, we show the intensity extracted along the red circular curve in Fig.~\ref{fig6} obtained for monolayer MoSe$_2$ together with the fit result using the formula described in Supplementary Information Note 4. The observed angle dependence matches those obtained by polarization-resolved SHG (Fig.~\ref{fig5}a) and ellipticity measurements (Fig.~\ref{fig5}b) discussed above, which were recorded for the same flake and orientation.

Additionally, we find a similar 6-fold pattern in the real-space SHG image, though it exhibits a strong dependence on defocusing, as demonstrated in Supplementary  Information Note 5. For practical purposes, 
Fourier SHG patterns are far more useful because they retain their symmetry properties and orientation even under significant defocus conditions.
The signal contribution in the center of the experimental patterns in Fig.~\ref{fig4} d,g,h, and Fig.~\ref{fig6} indicate incomplete destructive interference of SHG generated at different positions within the focus, which could be due to deviations from pure azimuthal laser polarization, sample roughness
and crystallographic disorder.

In addition to MoSe$_2$, Fig.~\ref{fig6} presents azimuthal mode Fourier SHG images of WSe$_2$ and h-BN, which exhibit the same 6-fold patterns, reflecting their D$_{3h}$ symmetry and respective orientations. 
A detailed sequence of measurements recording the 6-fold pattern using polarisation-resolved SHG and the Fourier SHG ellipticities for the samples shown in Fig.~\ref{fig6} can be found in Supplementary Information Video 1. 
The presented data thus confirms the general applicability of the presented approach to other materials with D$_{3h}$ symmetry.

\begin{figure}[htp]
\begin{center}
 \includegraphics[width=1\columnwidth]{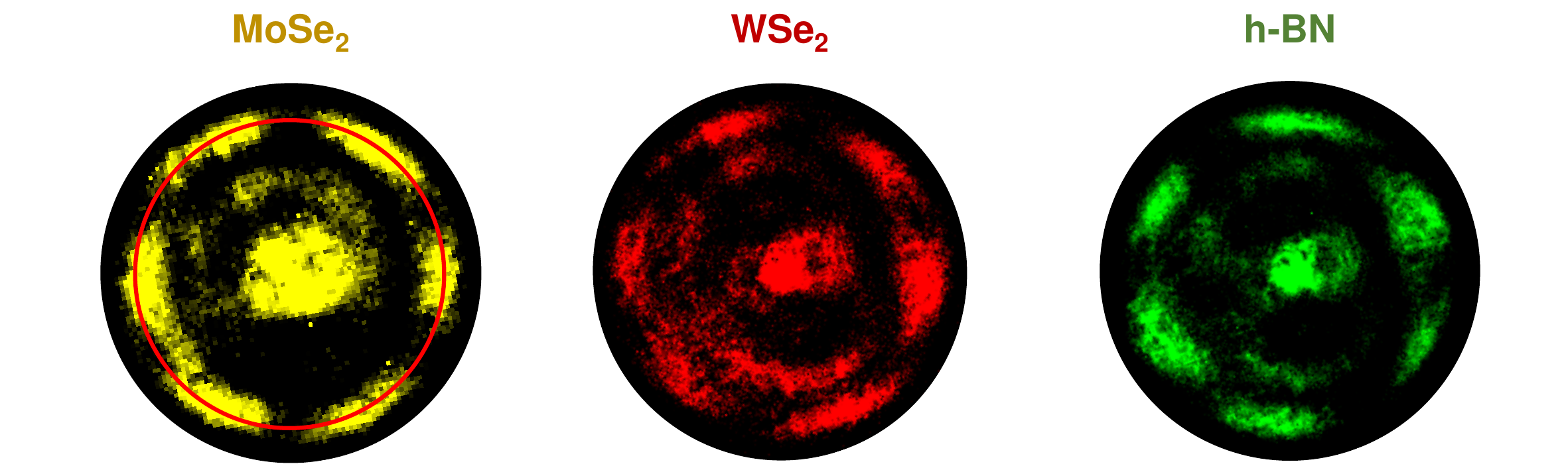}
\end{center}
 \caption{{\bf Azimuthal mode Fourier SHG images of different $D_{3h}$ materials.} 
 Fourier SHG pattern using the azimuthal mode for MoSe$_2$ (yellow), WSe$_2$ (red), and h-BN (green). A series of measurements for the samples presented in {\bf d} is included in  .
}
\label{fig6} 
\end{figure}

\section{CONCLUSION}

In summary, we demonstrated that Fourier space SHG imaging of noncentrosymmetric 2D materials enables the retrieval of information about the symmetry and orientation of their crystal lattice.
We first investigated the SHG radiation patterns obtained for a tightly focused Gaussian laser beam. We found a complex interplay between the excitation field's elliptical distribution and the radiation pattern's ellipticity created by a coherent superposition of dipolar SHG emitters in the focus generating the second harmonic light. 
The resulting ellipticity of the SHG Fourier patterns allows for the retrieval of the crystal orientation and provides a test case for the developed microscopic model that describes SHG image formation. However, the underlying procedure is experimentally demanding and, therefore, less practical.

We proposed and implemented a more direct approach using an azimuthally polarized laser beam, which encodes polarization information in the spatial coordinates of the excitation field. 
The recorded Fourier SHG images directly reflect the crystal lattice 
symmetry and its orientation, which can be described quantitatively by our model calculations.
We further note that the approach could also be used to investigate other noncentrosymmetric materials by explicitly considering the crystal symmetry through the associated second-order nonlinear susceptibility tensor.
The presented approach provides a rapid and precise tool that could readily be incorporated into nanofabrication procedures for the assembly of twisted 2D material and materials characterization in general.  

\section{METHODS}

{\bf SHG imaging map and BFP detection}\\
 Experiments were performed on the confocal microscope setup illustrated in Fig.~\ref{fig1} a and in Supplementary Information Note 1. For SHG and 2PL experiments, we use a femtosecond pulsed laser source operating at 880 nm \cite{birkmeier_wavelength-tunable_2024}. Silver mirrors guided the beam to an inverted microscope after being reflected by a non-polarising beamsplitter and focused by a NA=1.3 immersion oil objective onto the sample. Generated SHG and 2PL were collected by the same objective and spectrally filtered using a short pass and a band pass filter centered at 440 nm in the case of SHG, and an 850 nm short pass and a 600 nm long pass filter for 2PL. BFP patterns were measured using a CCD camera. Fourier and real space imaging modes are switched by exchanging the final lens with a lens with half the focal length. The samples are mapped on a scanning unit while the light is detected by a sensitive avalanche photodiode (APD). To record polarization-dependent SHG intensities, we rotated the incident Gaussian beam with a half-wave plate in the input beam path parallel to a polarizer in the detection beam path. Fourier space SHG images are detected without a polarizer in the detection beam path. 
 The azimuthal laser beam was generated using a laser mode converter. 

{\bf Preparation of 2D materials}

Using commercial bulk crystals, we performed mechanical exfoliation~ \cite{methods_materials_1} onto polydimethylsiloxane (PDMS) stamp to thin down the TMD multi-flakes. Monolayers were identified by optical contrast imaging. 
The selected monolayers were transferred onto 160 $\mu$m glass cover slides at 70 °C by the dry-transfer method~\cite{Methods_2_Castellanos-Gomez_2014}. 
Bright field contrast images, photoluminescence spectra and confocal SHG maps are included in Supplementary Information Note 2. 

{\bf Modelling of SHG images}
The incident field distributions of focused laser beams can be calculated analytically as described, for example, by Hecht \textit{et al.} \cite{hecht_propagation_2012}. In the calculations, the incident field is first projected onto the crystal axis, where the x-direction is taken to be parallel to a zigzag direction of the crystal lattice  (Fig.~\ref{fig1}b).
The induced polarization at the second harmonic $P=(P_x,P_y,0)$ is then calculated as $P_x=2 E_x E_y$ and  $P_y=E_x^2-E_y^2$
where $E_x$  ($E_y$) is the incident electrical field component parallel (perpendicular) to the zigzag (armchair) direction. 
Due to symmetry reasons, only three components of the second-order susceptibility tensor $\chi^{(2)}$ are nonzero with in-plane polarization.
As can be seen for incident field polarization parallel to the zigzag axis, the SHG polarization is rotated by 90°. In contrast, the polarization direction 
remains unaffected for incident polarization parallel to the armchair direction.
The Fourier SHG radiation pattern detected in the back-focal plane of the microscope objective is then calculated as the coherent sum of the radiation patterns generated by point dipoles as described in ref.~\citenum{lieb_single-molecule_2004} in the
illuminated sample area weighted by the local field distribution ($E_x(x,y)$, $E_y(x,y)$) of the strongly focused laser beam. A propagation factor of 
$f(x,y,\phi,\theta)=\exp(-i k_1 (x \cos(\phi) \sin(\theta)+ y \sin(\phi) \sin(\theta))$ accounts for retardation. Here, $k_1$ denotes the wavevector in the glass substrate, $\phi$ the in-plane orientation of the emitting dipole, and $\theta$ the angle of the emitted wave relative to the optical axis.
From the Fourier space pattern, the corresponding real space image is calculated using the integral forms in ref.~\citenum{hecht_propagation_2012} (Supplementary Information Note 3).

\renewcommand{\refname}{REFERENCES}

\bibliography{ArXiv_Main_paper}

\section{ACKNOWLEDGEMENTS}
L.~L. and L.~M. acknowledge support from the Alexander von  Humboldt Foundation. 
We acknowledge financial support from the Deutsche
Forschungsgemeinschaft (DFG) through
Germany's Excellence Strategy-EXC 2089/1-390776260. 

\section{AUTHOR CONTRIBUTIONS}
 All authors wrote and reviewed the manuscript.

\section{COMPETING INTERESTS}
The authors declare no competing interests.

\section{SUPPLEMENTARY INFORMATION}
Supplementary information accompanies this paper.

\newpage

\end{document}



{\bf Table of content}

\begin{itemize}
\itemsep0pt
\item Supplementary Note 1: Experimental setup
\item Supplementary Note 2: Sample Characterization
\item Supplementary Note 3: Model for Fourier space SHG in 2D materials
\item Supplementary Note 4: Accuracy of crystal orientation determination
\item Supplementary Note 5: Focus dependence of SHG images for azimuthal laser mode
\item Supplementary References
\end{itemize}

\newpage
\noindent
{\bf Supplementary Note 1: Experimental setup}

\begin{figure}[ht!]
    \centering
    \includegraphics[width=0.5\columnwidth]{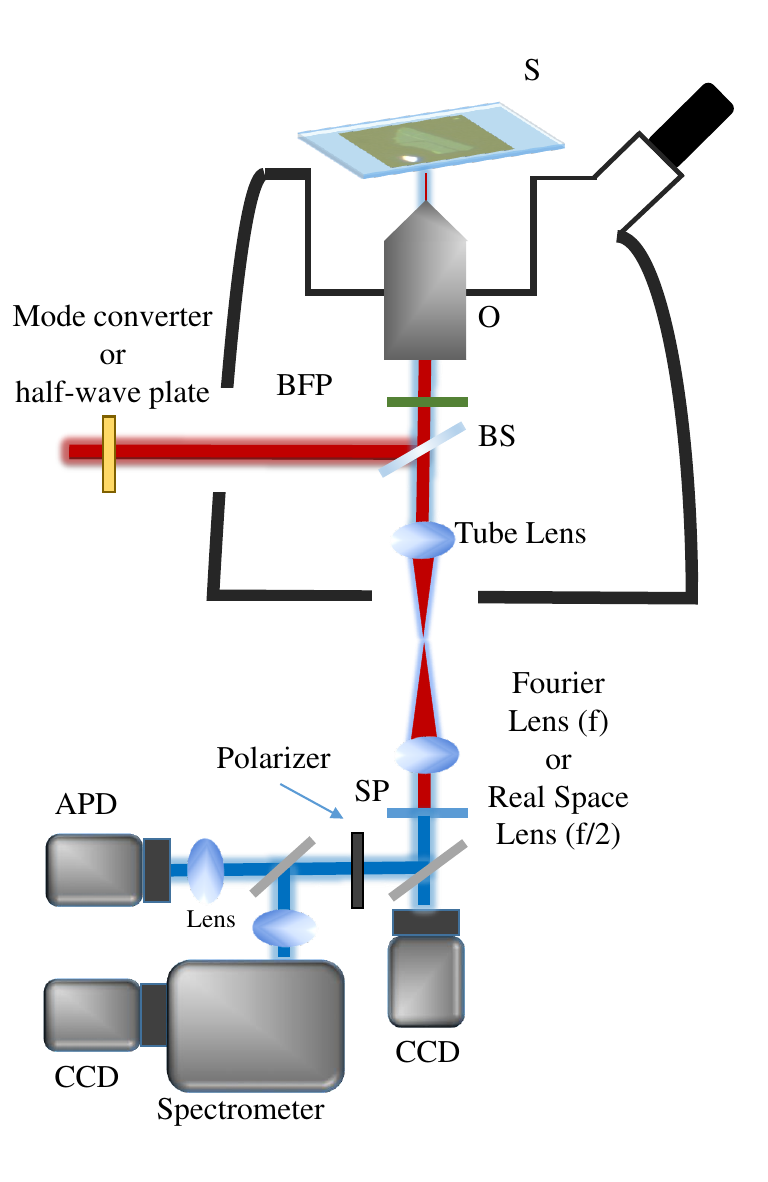}
    \caption{{\bf Illustration of the experimental setup:}
    The setup consists of a femtosecond pulsed source operating at 880 nm \cite{birkmeier_wavelength-tunable_2024}.
    The incident laser pulses (red) pass a half-wave plate (for Gaussian laser mode) or a mode converter (for azimuthal laser mode), are reflected by a beam splitter (BS) and focused by a microscope objective (O) (immersion oil objective, 100x, NA=1.3) on the sample (S). The sample response (SHG, 2PL) is detected in epi-configuration after passing the BS and a spectral short-pass filter (SP) to suppress scattered light at the laser wavelength. Fourier space images are recorded using a CCD camera in combination with a Fourier lens with focal length f positioned at a distance f from the focus of the tube lens and the CCD camera, respectively. Real space images are obtained by replacing the Fourier lens with a lens of f/2. A flip mirror is introduced to guide the light to a spectrometer equipped with a CCD or an avalanche photodiode (APD) for measuring spectra and confocal images. A polarizer is introduced to measure polarization resolved SHG.}
    \label{Sup_Figure_1}
\end{figure}

\pagebreak

{\bf Supplementary Note 2: Sample Characterization}

\begin{figure}[ht!]
    \centering
    \includegraphics[width=1\columnwidth]{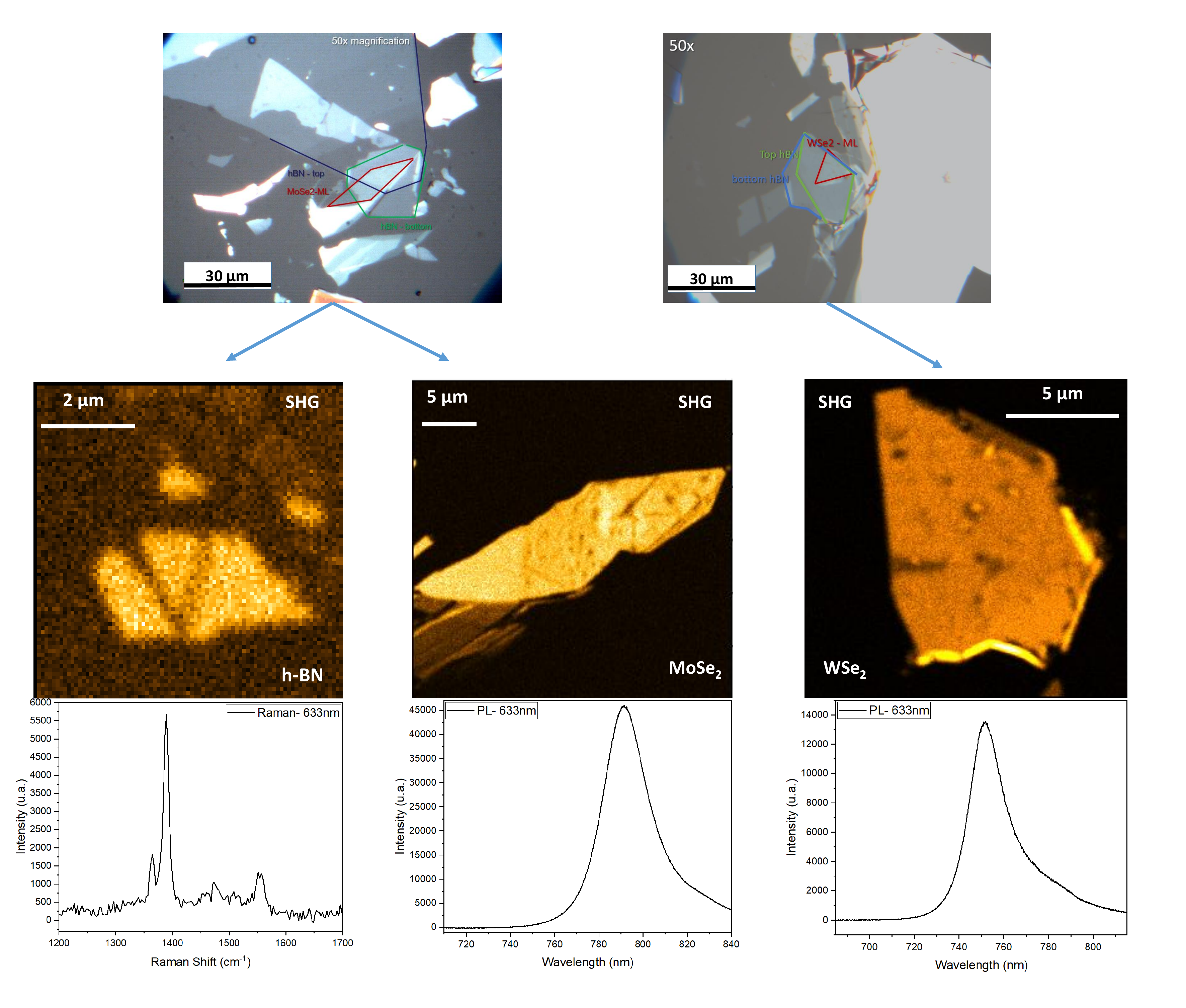}
    \caption{{\bf Sample characterization:} Top: Bright-field contrast images of the encapsulated samples: MoSe$_2$ together with h-BN (left )and WSe$_2$ (right). Middle: Confocal SHG images were recorded using the APD for h-BN (left), MoSe$_2$ (middle), and WSe$_2$ (right). Bottom: Raman spectrum of the h-BN sample showing the E$_{2g}$ mode characteristic of h-BN (left). Photoluminescence spectrum of MoSe$_2$ (middle) and WSe$_2$ (right) with the characteristic photoluminescence emission of monolayers.}
    \label{Sup_Figure_2}
\end{figure}

 All flakes were made by micromechanical exfoliation and encapsulated by transference using PDMS\cite{mennel_second_2018, Methods_2_Castellanos-Gomez_2014}. The samples were characterized using optical wide-field imaging and Raman spectroscopy for h-BN and photoluminescence (PL) for MoSe$_2$ and WSe$_2$. SHG confocal maps were recorded for all samples. The characterization results show monolayers for TMDs and a few layers for h-BN, as illustrated in Figure~\ref{Sup_Figure_2}.

\pagebreak

{\bf Supplementary Note 3: Model for Fourier space SHG in 2D materials}

\begin{figure}[ht!]
    \centering
    \includegraphics[width=0.5\columnwidth]{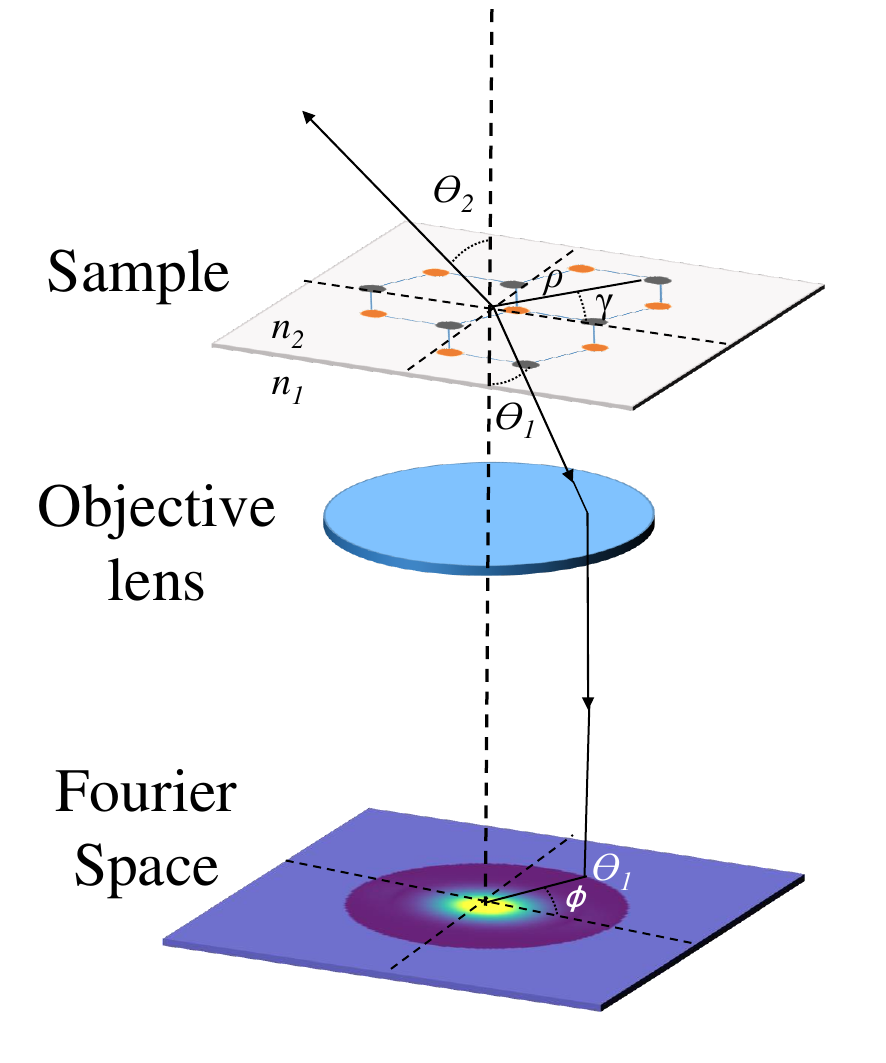}
    \caption{Illustration of the model used for calculating radiation patterns in Fourier space by back-focal plane (BFP) imaging.}
    \label{Sup_Figure_3}
\end{figure}

Radiation patterns in Fourier space and real space images are calculated from the sum of the fields generated by coherent dipoles within the 2D material layer oscillating at $2 \omega$ using a Phyton script. The strength of these dipoles is scaled with the local electric field of the exciting laser in the focus of the microscope objective~\cite{hecht_propagation_2012,lieb_single-molecule_2004}.
In the following, we define the parameters used in the available Phyton script.

\subsubsection{1. Parameters}
\begin{itemize}
     \item $n_1$ ($n_2$) Refractive index in the lower halfspace formed by the substrate (upper halfspace)
     \item $\theta_1$ ($\theta_2$) Collection angle in the lower halfspace (upper halfspace)
    \item $\lambda_{ex} (lex)$ Wavelength of the excitation light
    \item $\lambda (l)$ Wavelength of the SH light ($k_i=2 \pi n_i/ \lambda$)
     \item $NA$ Numerical aperture of the microscope objective 
     \item $ff$ Filling factor of the microscope objective
     \item $\phi$ Polar angle in the BFP
     \item $\gamma$ Polar angle in the real space focal plane ($\gamma=arctan(y,x)$), $\rho$ radius ($\rho=\sqrt{x^2+y^2}$)
\end{itemize}

\subsubsection{2. Fresnel coefficients for air-TMD-glass interface}
\begin{equation}
    r_p = \frac{n_1 \cos{\theta_2}-n_2 \cos{\theta_1}+Z_0 \sigma \cos{\theta_1} cos{\theta_2}}{n_1 \cos{\theta_2}-n_2 \cos{\theta_1}+Z_0 \sigma \cos{\theta_1} \cos{\theta_2}}, t_p=(1-r_p) \frac{\cos{\theta_2}}{\cos{\theta_1}}
\end{equation}
\begin{equation}
    r_s = \frac{n_2 \cos{\theta_2}-n_1 \cos{\theta_1}-Z_0 \sigma}{n_2 \cos{\theta_2}+n_1 \cos{\theta_1}+Z_0 \sigma}, t_s=r_s+1
\end{equation}

\subsubsection{3. Calculation of focal excitation field ~\cite{hecht_propagation_2012}}
Apodization function:
\begin{equation}
    f_w(\theta_1) = \exp{(-n_1/ff \sin{\theta_1}/NA)}
\end{equation}

Angular integrals:
\begin{equation}
    I_{00} = \int{f_w(\theta_1) \sqrt{\cos{\theta_1}}\sin{\theta_1}(1+\cos{\theta_1}) J_0(k_1 \rho \sin{\theta_1})d\theta_1}
\end{equation}
\begin{equation}
    I_{02} = \int{f_w(\theta_1) \sqrt{\cos{\theta_1}}\sin{\theta_1}(1-\cos{\theta_1}) J_2(k_1 \rho \sin{\theta_1})d\theta_1} 
\end{equation}
\begin{equation}
    I_{11} = \int{f_w(\theta_1) \sqrt{\cos{\theta_1}}\sin^2{\theta_1}(1+3\cos{\theta_1}) J_1(k_1 \rho \sin{\theta_1})d\theta_1} 
\end{equation}
\begin{equation}
    I_{12} = \int{f_w(\theta_1) \sqrt{\cos{\theta_1}}\sin^2{\theta_1}(1-\cos{\theta_1}) J_1(k_1 \rho \sin{\theta_1})d\theta_1} 
\end{equation}

In-plane focal field components, oriented along $\alpha$ with respect to crystal axis:
\begin{enumerate}
\item Linearly polarized Gaussian mode
\begin{equation}
E_{x}(\rho,\gamma)=I_{00}(\rho)+I_{02}(\rho) \cos{(2 (\gamma-\alpha))},
E_{y}(\rho,\gamma)=I_{02}(\rho) \sin{(2 (\gamma-\alpha))}
\end{equation}
\item Azimuthally polarized doughnut mode
\begin{equation}
E_{x}(\rho,\gamma)=i (I_{11}(\rho)+3 I_{12}(\rho)) \sin{(\gamma-\alpha)},
E_{y}(\rho,\gamma)=-i (I_{11}(\rho)+3 I_{12}(\rho))\cos{(\gamma-\alpha)}
\end{equation}
\end{enumerate}

Projected on to crystal axis:
\begin{equation}
E_{x,c}(\rho,\gamma)=\cos{\alpha} E_x - \sin{\alpha} E_y, 
E_{y,c}(\rho,\gamma)=\sin{\alpha} E_x + \cos{\alpha} E_y
\end{equation}

\subsubsection{4. SHG}
\begin{equation}
P_{x,c}^{2\omega}(\rho,\gamma)= 2 E_{x,c} E_{y,c}, 
P_{y,c}^{2\omega}(\rho,\gamma)= E_{x,c}^2 - E_{y,c}^2
\end{equation}

\subsubsection{5. Radiation pattern ~\cite{hecht_propagation_2012}}

Emission coefficients:
\begin{equation}
c_2= \frac{n_1}{n_2} t_p,  c_3= - \frac{n_1}{n_2} \frac{\cos{\theta_1}}{\cos{\theta_2}} t_s
\end{equation}

p- and s-polarized fields emitted by single point dipole:
\begin{equation}
E_p^{2\omega}(\rho,\gamma;\theta_1,\phi)= c_2 \cos{\theta_1} (\cos{\phi} P_{x,c}+\sin{\phi} P_{y,c}), 
E_s^{2\omega}(\rho,\gamma;\theta_1,\phi)= c_3 (\sin{\phi} P_{x,c}-\cos{\phi} P_{y,c}) 
\end{equation}

Components in x- and y-direction:
\begin{equation}
E_x^{2\omega}(\rho,\gamma;\theta_1,\phi)=\cos{\phi} E_p - \sin{\phi} E_s, 
E_y^{2\omega}(\rho,\gamma;\theta_1,\phi)=\sin{\phi} E_p + \cos{\phi} E_s
\end{equation}

Propagation phase factor for dipoles at $(x,y)$
\begin{equation}
    p_f(x,y;\theta_1,\phi)= \exp{(-i k_1 (x \cos{\phi}\sin{\theta_1}+y \sin{\phi}\sin{\theta_1}))} 
\end{equation}

Total field in the BFP from the coherent sum of dipole field :
\begin{equation}
    E_{x,total}^{2\omega}(\theta_1,\phi) = \int{E_x^{2\omega}(x,y;\theta_1,\phi) p_f dx dy},     E_{y,total}^{2\omega}(\theta_1,\phi) = \int{E_y^{2\omega}(x,y;\theta_1,\phi) p_f dx dy}
\end{equation}

Detected Fourier SHG patterns in BFP with $k_x/k_0= n_1 \sin{\theta_1} \cos{\phi}$ and $k_y/k_0= n_1 \sin{\theta_1} \sin{\phi}$:
\begin{equation}
    I^{2\omega}(\theta_1,\phi) = \frac{1}{\cos{\theta_1}} (| E_{x,total}^{2\omega}|^2+| E_{y,total}^{2\omega} |^2)
\end{equation}

2PL Fourier patterns are calculated from the incoherent sum of the intensity BFP patterns of two orthogonal in-plane dipoles with equal strength~\cite{budde_raman_2016}. 

Real space images are calculated by further propagating these fields followed by focusing~\cite{hecht_propagation_2012}.

\pagebreak

{\bf Supplementary Note 4: Accuracy of crystal orientation determination}

\begin{figure}[ht!]
    \centering
    \includegraphics[width=0.9\columnwidth]{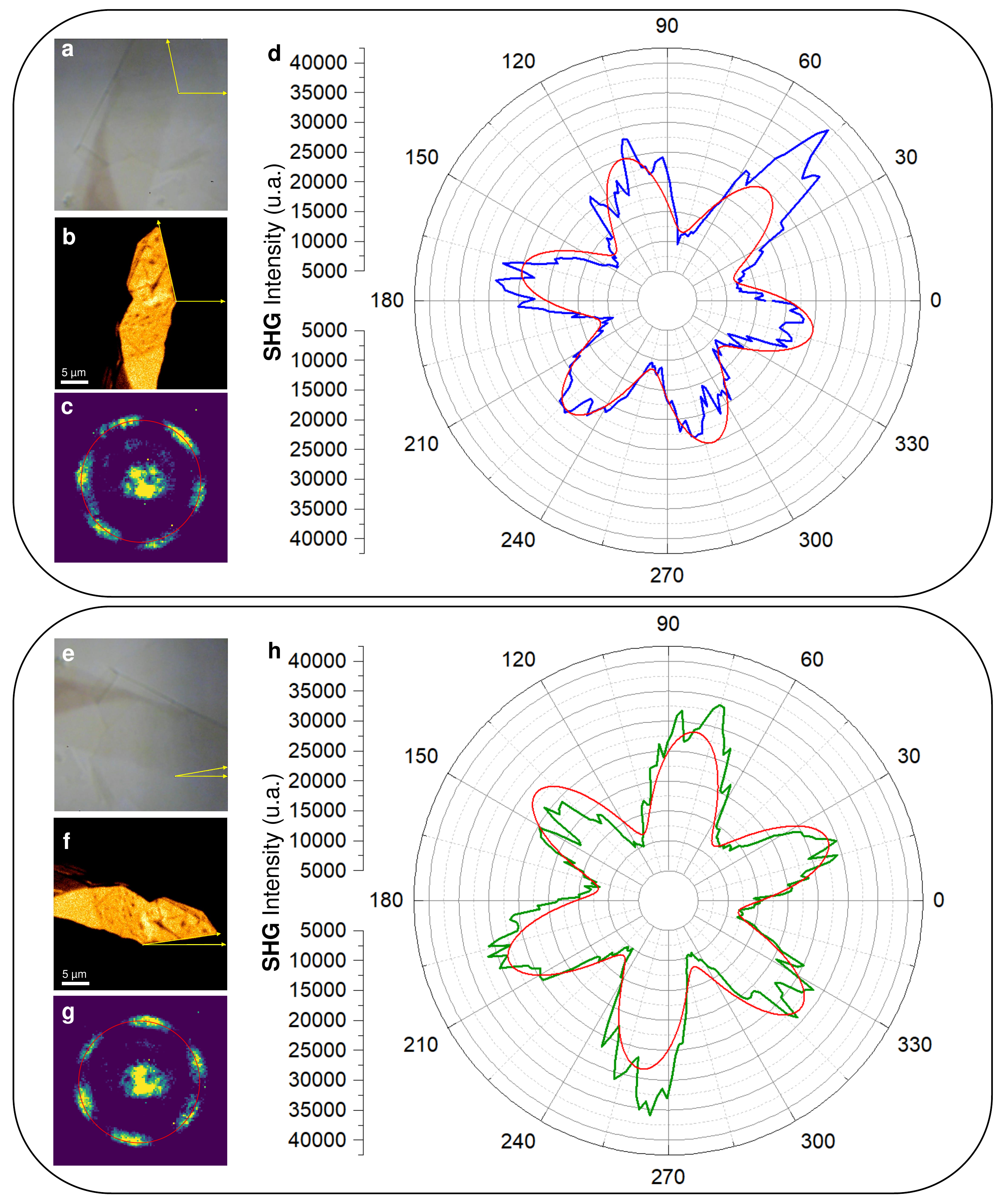}
    \caption{Measurement of a MoSe$_2$ monolayer in two different in-plane orientations. {\bf a,e} Wide-field optical images with the orientational angle determined by ImageJ.{\bf b,f} Confocal SHG maps measured using the APD with the angles determined by ImageJ. {\bf c, g} Fourier space SHG images using the azimuthal mode and {\bf d,h} polar plots together with fits of the red circle from image {\bf c, g} using eqn.\ref{eq:angledep}.}
    \label{Sup_Figure_4}
\end{figure}

We determined the accuracy of the crystal orientation determination using Fourier SHG imaging for azimuthal mode excitation after rotating the sample by different angles and comparing the derived values with those observed in real space optical images. In Fig.~\ref{Sup_Figure_4}, we recorded wide-field optical images (a, e) and the corresponding confocal SHG maps (b, f). We measured the angle between the indicated sample edge relative to 0° (horizontal yellow line). The angles were determined using the ImageJ software (see Table 1). Afterwards, we detected the Fourier SHG images using the azimuthal mode (c, g). Then, we made a polar plot of the SHG intensity along the red circle in the images in Fig.~\ref{Sup_Figure_4} c, g. These plots are shown in Fig.~\ref{Sup_Figure_4} d, h, and were fitted (red curves in d, h) using the following equation \cite{malard_observation_2013}:

\begin{equation}
    I^{SHG}(\phi) = (A\sin{3(\phi+P)})^2+C
    \label{eq:angledep}
\end{equation}

Here, A is the SHG amplitude, P represents the angular orientation, and C is a small uniform background. 
The relative angular orientation extracted from the fit demonstrates a higher accuracy than the direct determination using ImageJ (see Fig.~\ref{Sup_Figure_4}). In Table 1 below, we show the angles extracted from ImageJ for the sample of MoSe$_2$ measured by wide-field illumination, the same sample under the same conditions mapped by confocal SHG imaging, and the values extracted from the fit to the polar plot of the Fourier SHG data. The uncertainty of this fit is $\pm 0.5$ as given by its standard deviation. The last row lists the angular difference between the sample in position Fig.~\ref{Sup_Figure_4} a and Fig.~\ref{Sup_Figure_4} e (a-e) determined by the different techniques.

\begin{center}
{\bf Table 1}: Orientation angles obtained from different techniques for MoSe$_2$.
\begin{tabular}{||c| c c c||} 
 \hline
 MoSe$_2$ & Wide-field & SHG Map & Fourier SHG azimuthal mode\\ [0.5ex] 
 \hline\hline
 a, b, c & 102.0° & 102.3° & 102.2° \\ 
 \hline
 e, f, g & 9.9° & 8.3° & 9.2°  \\
 \hline \hline
 a - e & 92.1° & 94.0° & 93.0°  \\
 \hline
\end{tabular}
\end{center}

\pagebreak

\begin{figure}[ht!]
    \centering
    \includegraphics[width=.9\columnwidth]{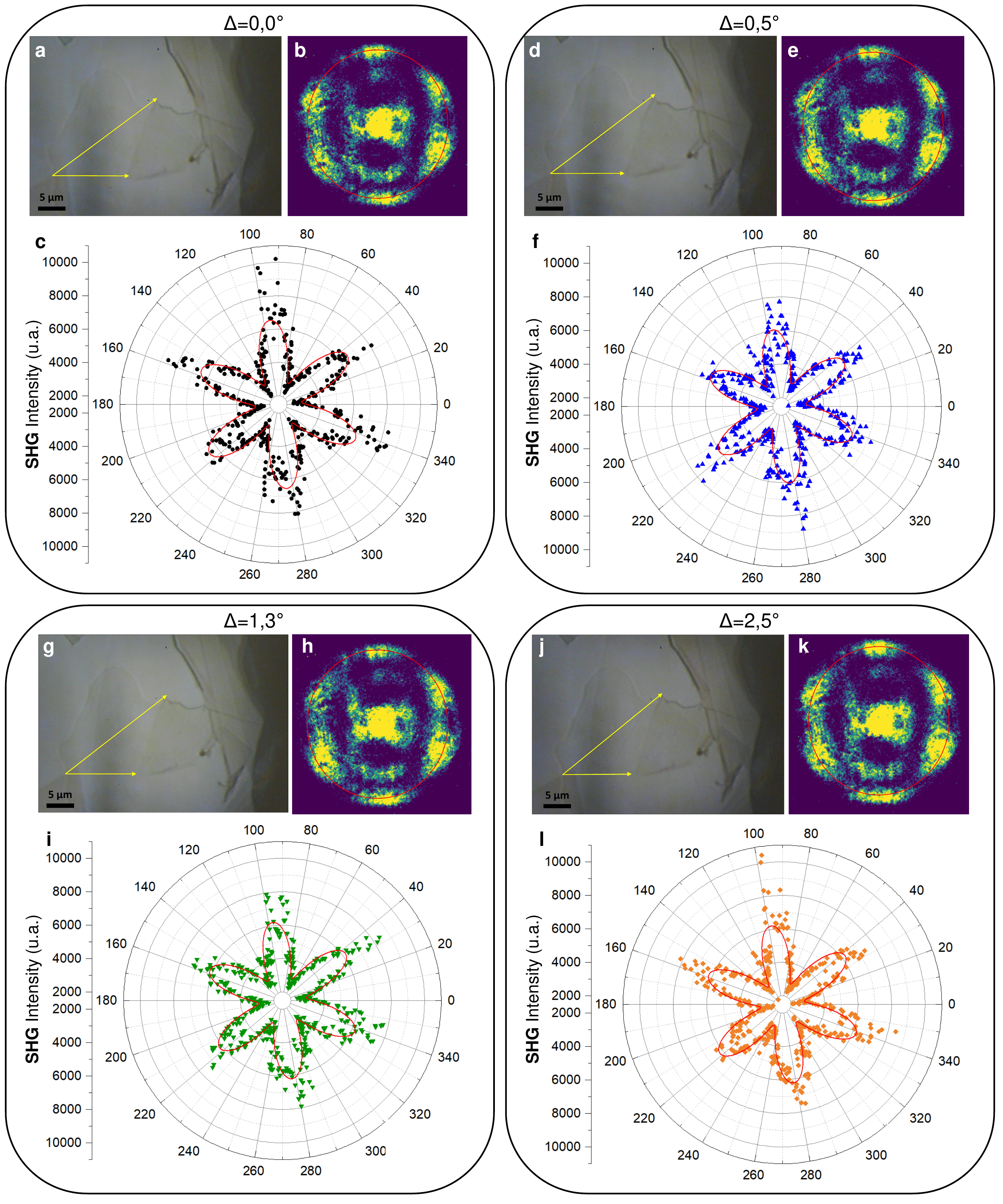}
    \caption{Measurement of a WSe$_2$ monolayer after rotation by small angles. {\bf a, d, g, j} wide-field images, the angle of the monolayer was determined by ImageJ. {\bf b, e, h, k}, Fourier SHG imaging was done using the azimuthal mode. {\bf c, f, i, l} polar plots of the red circles from {\bf b, e, h, k} together with the fits using eqn.~\ref{eq:angledep}.}
    \label{Sup_Figure_5}
\end{figure}

To confirm the accuracy of the Fourier SHG technique, we measured a different sample (WSe$_2$) rotated by small angles, as shown in Fig.~\ref{Sup_Figure_5}. The values obtained in the experiment of Fig.~\ref{Sup_Figure_5} are presented in Table 2. The table lists the changes in angular orientation $\Delta$ relative to the first orientation seen in Fig.~\ref{Sup_Figure_5} a, b, c.

\begin{center}
{\bf Table 2}:  Changes in orientation angle obtained from different techniques for WSe$_2$.

\begin{tabular}{||c| c c ||} 
 \hline
 MoSe$_2$ & $\Delta$ (Wide-field) & $\Delta$ (Fourier SHG azimuthal mode)  \\ [0.5ex] 
 \hline\hline
 a, b, c &  0.0° & 0.0°\\ 
 \hline
 d, e, f& 0.6° & 0.5° \\
 \hline
 g, h, i & 1.3° & 1.3°  \\
 \hline
 j, k, l & 2.3° & 2.5° \\
 \hline
\end{tabular}
\end{center}

\pagebreak

{\bf Supplementary Note 5: Focus dependence of SHG images for azimuthal laser mode}

\begin{figure}[ht!]
    \centering
    \includegraphics[width=1\columnwidth]{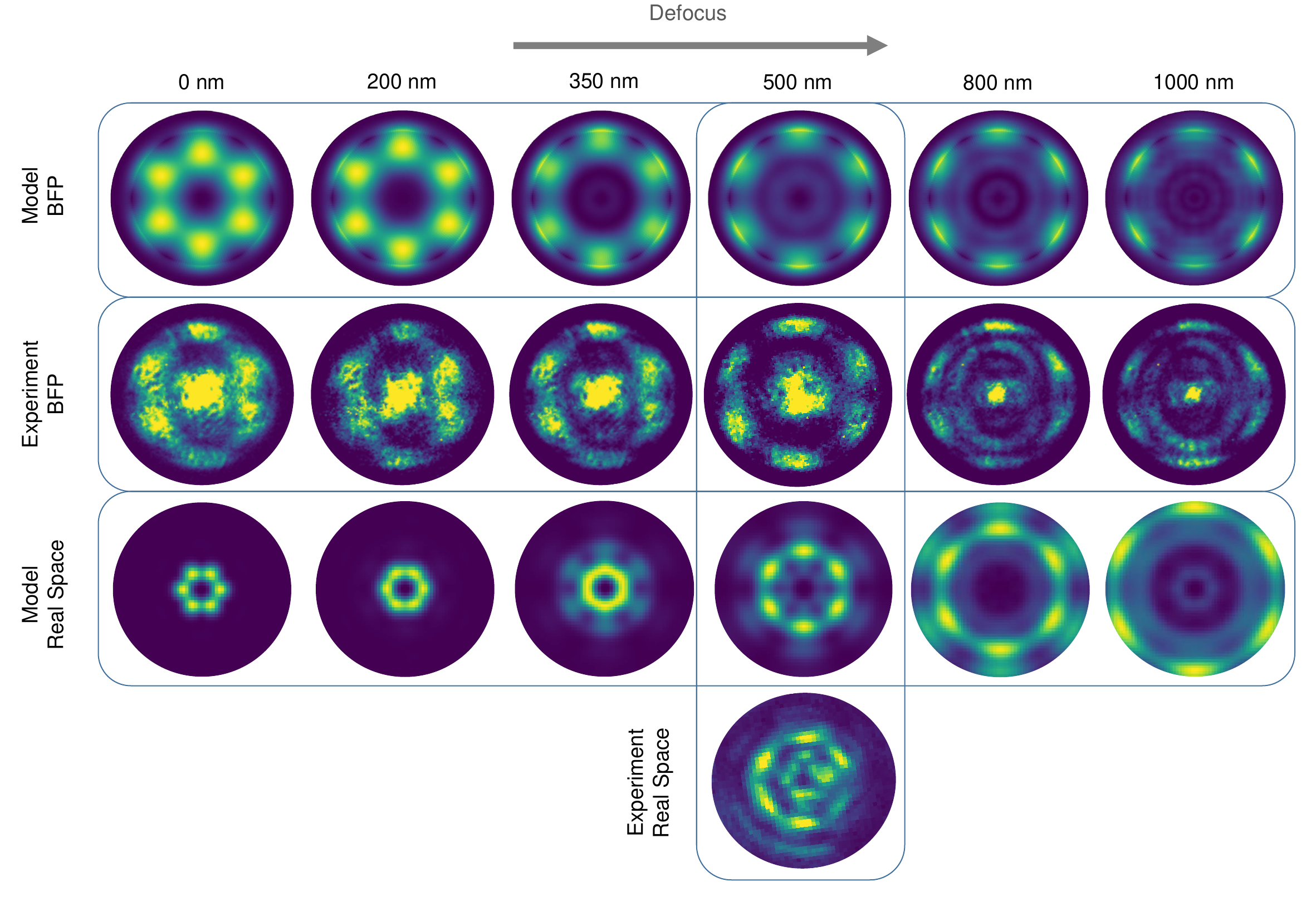}
    \caption{Focus dependence of calculated and experimental SHG images of MoSe$_2$ for an azimuthally polarized laser mode. Upper two rows: Fourier space BFP patterns. Lower two rows: Corresponding real space patterns.}
    \label{Sup_Figure_6}
\end{figure}

Using the model described in the main manuscript and in Supplementary Note 3, we can simulate Fourier and real space images for the azimuthally polarized laser mode for different defocus conditions. 
Upon defocusing, the Fourier SHG images retain their 6-fold symmetry and orientation (Fig.~\ref{Sup_Figure_6}), as can be seen in both calculated and experimental patterns.
Whereas the real space images always display 6-fold patterns, their orientation changes upon defocusing.
In focus, the real space image shows a hexagonal pattern that is rotated with respect to the Fourier space pattern by 30°.
Fourier space SHG detection thus provides a robust way to visualize the crystal orientation.

\pagebreak

\renewcommand{\refname}{REFERENCES}

\bibliography{Supplementary_information}